\documentclass[12pt]{article}

\usepackage[letterpaper, margin=1.2in, noheadfoot]{geometry}
\usepackage{amsmath}
\usepackage{amssymb}
\usepackage{graphicx}
\usepackage{natbib}
\usepackage{color}

\usepackage{url}

\newcommand{\s}{s}
\newcommand{\var}{\sigma}
\newcommand{\pd}{\partial}

\begin{document}
\title{Statistical distributions of pyrosequencing}
\author{Yong Kong\\
Department of Molecular
Biophysics and Biochemistry\\
W.M. Keck Foundation Biotechnology Resource Laboratory \\
Yale University\\
333 Cedar Street, New Haven, CT 06510\\
email: \texttt{yong.kong@yale.edu} }

\date{}

\maketitle

\newpage

\begin{abstract}
Pyrosequencing is emerging as one of the 
important next-generation sequencing technologies.
We derive the statistical distributions
of this technique in terms of nucleotide probabilities
of the target sequences.
We give exact distributions both for fixed number of flow cycles
and for fixed sequence length.
Explicit formulas are derived for the mean and variance
of these distributions.
In both cases, the distributions can be approximated accurately
by normal distributions with the same mean and variance.
The statistical distributions will be useful 
for instrument and software development for pyrosequencing
platforms.
\end{abstract}

\newpage

\section{Introduction} \label{S:intro}
The emerging new sequencing platforms, 
the so-called next-generation sequencing technology,
have enabled researchers to generate more sequencing data than ever before
with a dramatically reduced cost.
The innovations have already transformed the way biology experiments are
carried out.
Unlike the traditional Sanger capillary electrophoresis method
using dideoxynucleotide chain-termination,
many of these new platforms explore the concept of 
sequencing by synthesis.
Pyrosequencing technique is one of such approaches 
\citep{Ronaghi1998}.  
Compared with other next-generation sequencing techniques,
currently the pyrosequencing technology has the advantage of 
longer sequence read length,
which makes it possible for \emph{de novo} sequencing of new genomes.

In this paper we derive the statistical distributions
for the pyrosequencing technique.
These distributions are useful for various stages
in the use and development of the pyrosequencing technology,
such as
instrument development and testing, 
algorithm and software development,
and the everyday machine performance monitoring and trouble-shooting.

The distribution of the number of flow cycles for sequences with
fixed length 
and the distribution of sequence length at fixed number of flow cycles
are obtained.  In both cases the distributions can be approximated 
quite accurately
by normal  distributions
with the same mean and variance as the exact distributions.
We obtained the distributions by using the method of probability
generating functions (GFs), which in turn were obtained by using 
the recurrence relations between the probabilities
at different sequence lengths and flow cycles.

The paper is organized as follows.
First in the remaining of this \emph{Introduction} section
 we give a brief description of the pyrosequencing technique,
which is useful for the subsequent theoretical developments.
We also define the necessary notation here.
The derivation of the main results, 
which are exact under the assumptions of the sequence model,
will be presented in the \emph{Bivariate Generating Functions} section.
After that, we present the explicit 
formulas for the mean and variance of the distributions,
for both fixed number of cycles and fixed sequences length.
We show that the exact distributions can be approximated
accurately by normal distributions 
with the same mean and variance calculated from these formulas.
These explicit formulas would be most useful for the practitioners
in the field.
We also present the results for individual flows (the four different
nucleotides) in this section.

\subsection{Pyrosequencing technique}
Pyrosequencing protocol is based on the detection of the pyrophosphate (PPi)
that is released during DNA synthesis.
In most of the current systems, several enzymes are involved in the chemical
reactions.
The protocol adds the four kinds of nucleotides (dATP, dCTP, dGTP, and dTTP) 
stepwise and iteratively, 
with one kind of nucleotide at a time.
For each nucleotide flow, 
if the added nucleotide is complementary to the
DNA template being sequenced, the added nucleotide is incorporated
by polymerase.
Inorganic PPi is released as a result of the polymerization reaction,
which is converted to ATP by ATP sulfurylase. 
The ATP provides energy for luciferase to oxidize luciferin
and to generate detectable light quantitatively. 
Because we know which nucleotide is added at each nucleotide flow, 
the DNA sequence of the
template can be determined by the presence or absence of the emitted light
and the intensity of the emitted light.
The detected light is usually presented as a \emph{pyrogram}, 
in which the x-axis is the pre-determined nucleotide flows
and y-axis is the intensity of the emitted light at each nucleotide flow. 
Excess nucleotides are enzymatically degraded 
before the next nucleotide is added.  
Ideally the intensity of the light is proportional to
the number of incorporated nucleotides.
In reality this is not always the case due to the nonlinear light response
following incorporation of more than a few identical
nucleotides.
This poses an inherent difficulty for the pyrosequencing protocol
to determine the homopolymeric region accurately.

\subsection{Notation and definitions}

To avoid the unnecessary specification of the
detailed names of the four kinds of nucleotides, in the following we will
use $a$, $b$, $c$, and $d$ to represent any permutations of 
the usual nucleotides $A$, $C$, $G$, and $T$.
Throughout the paper we assume that the nucleotides in the target
sequence are independent of each other. 
The probabilities for the four nucleotides in the target sequence
are denoted as 
$p_a$, $p_b$, $p_c$, and $p_d$. 
In Table~\ref{T:defs} we define the nucleotide \emph{flow cycle number} $f$
(the third row) to distinguish it from nucleotide flow number
(the second row).
A flow \emph{cycle}  
is the ``quad cycle'' of successive four nucleotides \{$abcd$\}.
The cycle number is denoted as $f$ in the following.
We will use $n$ for the length of a sequence. 

In Table~\ref{T:defs} we also show the \emph{ideal} pyrograms for
two hypothetical sequences with the same length of $9$:
\verb|bddabaaad| and \verb|abbbdabbc|.
Although the length of the two sequences is the same,
they need different number of flows (and hence number of cycles)
to determine their sequences, 
due to the particular arrangements of the nucleotides
in the sequences
in relation to the order of nucleotide flow: 
the first sequence needs $12$ flows ($3$ cycles),
while the second sequence needs $7$ flows ($2$ cycles).
In the following we will determine
the statistical distributions they will follow.
It is easy to see that at a fixed sequence length $n$,
the minimum number of nucleotide flows is $1$ for a stretch of $n$ $a$'s,
while the maximum number of nucleotide flows is $3n + 1$
for sequence \verb|dcbadcba...|.


\begin{table}
\caption{
The definitions of flow number and cycle number,
and the pyrograms for two sequences with the same length of $9$:
``bddabaaad'' (pyrogram-1)  and ``abbbdabbc'' (pyrogram-2).
}
\label{T:defs}
\begin{tabular}{lcccccccccccc}
\hline\\
nucleotide flow        & a & b & c & d & a & b & c & d & a & b  & c  & d \\
flow number            & 1 & 2 & 3 & 4 & 5 & 6 & 7 & 8 & 9 & 10 & 11 & 12 \\
cycle number ($f$)     & 1 & 1 & 1 & 1 & 2 & 2 & 2 & 2 & 3 & 3 & 3 & 3 \\
pyrogram-1         & 0 & 1 & 0 & 2 & 1 & 1 & 0 & 0 & 3 & 0  & 0  & 1 \\
pyrogram-2         & 1 & 3 & 0 & 1 & 1 & 2 & 1 &   &   &    &    &
\end{tabular}
\end{table}

In the following we will frequently use the definition of
\emph{elementary symmetric functions} to express the results in compact forms.
For our purpose the elementary symmetric functions with four variables 
are defined in
Eq. (\ref{E:esf}), in terms of the nucleotide probabilities:
\begin{align} \label{E:esf}
  \s_1 &= p_a + p_b + p_c + p_d ,                                     \notag \\
  \s_2 &= p_a p_b + p_a p_c +  p_a p_d + p_b p_c + p_b p_d + p_c p_d ,\notag\\ 
  \s_3 &= p_a p_b p_c +  p_a p_b p_d + p_a p_c p_d + p_b p_c p_d ,    \notag\\
  \s_4 &= p_a p_b p_c p_d .                                           
\end{align}
Since there are only four nucleotides, apparently we have the constraint on
$s_1$ as 
$s_1 = p_a + p_b + p_c + p_d = 1$.

We'll frequently extract coefficients from the expansion of GFs.
If $f(x)$ is a series in powers of $x$, then we use the notation
$[x^n]f(x)$ to denote the coefficient of $x^n$ in the series.
Similarly, we use $[x^n y^m]f(x,y)$ to denote the 
coefficient of $x^n y^m$ in the bivariate $f(x, y)$.

\section{Bivariate Generating Functions} \label{S:exact}
In this section we first establish 
the recurrence relations between probabilities 
with respect to sequence length and flow cycle number,
and then derive probability GF from these recurrence relations.

\subsection{Recurrences}
Let $L_i(f, n)$, $i=a$, $b$, $c$, and $d$ denote 
the probability (up to a normalization
factor, see below) of sequences with
a length of $n$ that has a pyrogram of $f$ flow cycles
with the last nucleotide flow being $i$.
The following recurrence relations can be established:
\begin{align}
 L_a(f+1, n+1) &= \left[ L_a(f+1, n) + L_b(f, n) + L_c(f, n) + L_d(f, n)
                  \right] p_a ,      \notag  \\
 L_b(f+1, n+1) &= \left[ L_a(f+1, n) + L_b(f+1, n) + L_c(f, n) + L_d(f, n)
                  \right] p_b ,      \notag  \\
 L_c(f+1, n+1) &= \left[ L_a(f+1, n) + L_b(f+1, n) + L_c(f+1, n) + L_d(f, n)
                  \right] p_c ,       \notag  \\
 L_d(f, n+1) &= \left[ L_a(f, n) + L_b(f, n) + L_c(f, n) + L_d(f, n)
                  \right] p_d .
\end{align}
The recurrences cannot be solved in closed forms.
However, their GFs can be solved in compact forms.

\subsection{Generating functions}
The GFs of $L_i(f, n)$ are defined as
\begin{equation}
 G_i (x, y) 
 = \sum_{n=1}^\infty \sum_{f=1}^\infty L_i(f, n) x^f y^n,
 \qquad i = a, b, c, d.  
\end{equation}
By using proper initial conditions,
these GFs are solved as
\begin{align}
 G_a (x, y) &= \frac{p_a x y } {H} F                           , \label{E:G_a}\\
 G_b (x, y) &= \frac{p_b x y}{H}  
 \left[ 1 - (p_c + p_d) (1-x) y 
   + p_c p_d (1-x)^2 y^2
 \right]                        ,                 \label{E:G_b}\\
 G_c (x, y) &= \frac{p_c x y}{H} 
 \left[ 1 -  p_d (1-x) y\right] ,                 \label{E:G_c}\\
 G_d (x, y) &= \frac{p_d x y} {H}      ,                 \label{E:G_d}
\end{align}
where
\begin{multline}
 F = 
   [ 1 - (p_b + p_c + p_d) (1-x) y \\
     + (p_b p_c + p_b p_d + p_c p_d) (1-x)^2 y^2
     - p_b p_c p_d (1-x)^3 y^3
   ]
\end{multline}
and
\[
 H = 1 - y + \s_2 (1-x) y^2 - \s_3 (1-x)^2 y^3 + \s_4 (1-x)^3 y^4 .
\]
The $L_i(f, n)$ can be obtained from their corresponding GF $G_i(x, y)$
by extracting the appropriate coefficients.
By using the notation we introduced earlier
we have $L_i(f, n) = [x^fy^n] G_i (x, y)$, $i=a$, $b$, $c$, and $d$.

From the expressions of the GFs we can see that they are not symmetric
with respect to the nucleotide probabilities $p_i$, $i=a$, $b$, $c$, and $d$.
If we only consider the nucleotide flows  
that end up in the same ``quad cycle'' (see Table~\ref{T:defs}),
then we can add the four GFs together
to obtain
\begin{align} \label{E:G_sum}
G (x, y) &= G_a + G_b + G_c + G_d \notag \\
  &= \frac{xy}{H} [ 1 - s_2(1-x)y + s_3(1-x)^2 y^2 - s_4(1-x)^3 y^3] .
\end{align}
The expression of $G(x, y)$ is symmetric
with respect to the nucleotide probabilities, since all the 
parameters involved are encapsulated in $s_i$, $i=2,3,4$, 
the elementary symmetric functions of  the nucleotide probabilities.

\subsection{Normalization factors} \label{SS:norm}

If we set $x=1$ in these GFs, for example in $G_a(x, y)$,
we get $G_a(1, y) = \sum_{n=1}^\infty [ \sum_{f=1}^\infty L_a(f, n) ]  y^n$.
The inner sum in the bracket is the total sum of $L_a(f, n)$
over the full range of flow cycle $f$
for a given value of sequence length $n$.
This is 
the normalization factor for $L_a(f, n)$
when the sequence length is fixed at $n$.
From Eqs.~\ref{E:G_a}, \ref{E:G_b}, \ref{E:G_c}, and \ref{E:G_d}
we see that
\begin{align*}
 G_a(1, y) &= \frac{p_a y}{1 - y}, &  G_b(1, y) &=  \frac{p_b y}{1 - y}, \\
 G_c(1, y) &= \frac{p_c y}{1 - y}, &  G_d(1, y) &=  \frac{p_d y}{1 - y},
\end{align*}
which lead to 
the normalization factors
\begin{align} \label{E:normal_x}
  u_a &= \sum_{f=1}^\infty L_a(f, n) = [y^n]G_a(1, y) = p_a   ,  \notag \\ 
  u_b &= \sum_{f=1}^\infty L_b(f, n) = [y^n]G_b(1, y) = p_b   ,  \notag \\ 
  u_c &= \sum_{f=1}^\infty L_c(f, n) = [y^n]G_c(1, y) = p_c   ,  \notag \\ 
  u_d &= \sum_{f=1}^\infty L_d(f, n) = [y^n]G_d(1, y) = p_d   .  
\end{align}
Obviously the sum of these normalization factors equals to $1$,
which means that $G(x, y)$ in Eq.~(\ref{E:G_sum}) is a true probability GF 
at fixed sequence lengths.
If $L_a(f, n)$, $L_b(f, n)$, etc. 
are considered alone, they should be divided by 
their corresponding normalization factors
in order for them to be interpreted as true probability
at fixed sequence lengths.

Similarly, if we set $y=1$ in these GFs, 
we get $\sum_{f=1}^\infty [ \sum_{n=1}^\infty L_i(f, n) ]  x^f$.
The inner sum in the bracket is the total sum of $L_i(f, n)$
over the full range of sequence length $n$
for a given value of flow cycle $f$.
This is 
the normalization factor for $L_i(f, n)$
when the number of flow cycles is fixed at $f$.
It can be shown that these normalization factors are
\begin{align} \label{E:normal_y}
  v_a &= \sum_{n=1}^\infty L_a(f, n) = [x^f]G_a(x, 1) 
  \approx \frac{p_a}{s_2} , \notag \\ 
  v_b &= \sum_{n=1}^\infty L_b(f, n) = [x^f]G_b(x, 1) 
  \approx \frac{p_b}{s_2} , \notag \\ 
  v_c &= \sum_{n=1}^\infty L_c(f, n) = [x^f]G_c(x, 1) 
  \approx \frac{p_c}{s_2} , \notag \\ 
  v_d &= \sum_{n=1}^\infty L_d(f, n) = [x^f]G_d(x, 1) 
  \approx \frac{p_d}{s_2} . 
\end{align}
There are two extra terms in these expressions, but they are so small
for even moderate $f$ that 
practically they can be ignored: for the number of flow cycles
 as small as $f=10$,
the extra terms do not make any difference in the eleventh decimal place,
and the bigger the number of flow cycles $f$, the smaller contributions of these
extra terms.
For clarity reason, these small terms are not shown here.

By dividing these normalization factors, 
the $L_a(f, n)$, etc. can be interpreted as true probability
at the fixed number of flow cycles.
For the quad cycle, the normalization factor
 is the sum of the individual factors, which add up to $v \approx 1/s_2$. 

\subsection{Mean and variance}
The availability of GFs makes it easy to derive the mean and variance
for the exact distributions.
When the sequence length is fixed,
the mean and variance are given by
\begin{align}
 \bar{f} (n) &= [y^n]\frac{\pd G(x, y) } {\pd x} \Big |_{x=1}   , 
 \label{E:avg_fixed_length}\\
 \var^2_f (n)  &= [y^n]\frac{\pd^2 G(x, y) } {\pd x^2} \Big |_{x=1} 
 + \bar{f}(n) - \bar{f}^2(n) . \label{E:var_fixed_length}
\end{align}
Similar formulas apply to the individual nucleotide flow GFs $G_a$, $G_b$, 
etc,
with their corresponding normalization factors 
shown in Eq. (\ref{E:normal_x}).

When the number of flow cycles is fixed,
the mean and variance are given by
\begin{align}
 \bar{n} (f) &= s_2 [x^f]\frac{\pd G(x, y) } {\pd y} \Big |_{y=1}  ,  
 \label{E:avg_fixed_cycle}\\
 \var^2_n (f)  &= s_2 [x^f]\frac{\pd^2 G(x, y)} {\pd y^2} \Big |_{y=1} 
 + \bar{n}(f) - \bar{n}^2(f) .  \label{E:var_fixed_cycle}
\end{align}
Similar formulas apply to the individual nucleotide flow GFs,
with their corresponding normalization factors shown in Eq. (\ref{E:normal_y}).

\subsection{A note on numerical calculations}
It should be pointed out that all the numerical 
calculation carried out in the following
used exact calculation throughout without losing precision.
In other words, all coefficients in the expansion of the GFs are
either in integers or exact fractions. 
The expansion was done using PARI/GP, 
a computer algebra system~\citep{PARI2}.

If floating points were used, then for longer sequence length
or flow cycles, very high precisions would be needed in order to
guarantee accuracy.

\section{Distributions at fixed sequence length and 
fixed number of flow cycles}
We discuss our results in two different scenarios.
In the first case,
the length of the target sequences is fixed.
As shown in Table~\ref{T:defs},
the number of flow cycles that have to be consumed 
in order to determine the sequences of the same length
fluctuates from sequence to sequence.
The distribution of the number of flow cycles
will be discussed in the first section~\ref{ss:fixed_length} below.
In the second case,
the number of flow cycles is fixed.
In this scenario, the length of target sequences
that can be determined by the flow cycles
follows a different statistical distribution,
which will be discussed in section~\ref{ss:fixed_cycle}.

\subsection{Fixed sequence length: distribution of flow cycles} 
\label{ss:fixed_length}
When the length $n$ of the target sequences is fixed,
the mean ${\bar f} (n)$ 
and variance $\var_f^2 (n)$ of the number of flow cycles $f$ that is needed
to determine the sequences
can be calculated
from the exact probability GFs 
described in the previous section 
by using Eqs.~\ref{E:avg_fixed_length} and \ref{E:var_fixed_length} as:
\begin{align} 
 {\bar f} (n) &= \s_2 n - \s_2 + 1 \label{E:fixed_length_avg} ,\\ 
 \var_f^2 (n) &= (\s_2 - 3 \s_2^2 + 2 \s_3) n + (5 \s_2^2 - \s_2 - 4 \s_3) .
 \label{E:fixed_length_var}
\end{align}
From Eqs. (\ref{E:fixed_length_avg}) and (\ref{E:fixed_length_var}) we can see
that both the mean ${\bar f} (n)$ and the variance $\var_f^2 (n)$
increase linearly with the sequence length $n$.
For the special case when $p_a = p_b = p_c = p_d = 1/4$, we have
\begin{align} 
  {\bar f} (n) &= \frac{3}{8} n + \frac{5}{8} \label{E:fixed_length_eq_avg} ,\\
  \var_f^2 (n)     &= \frac{5}{64} n + \frac{5}{64}  .
  \label{E:fixed_length_eq_var}
\end{align}
By using the constraint $s_1 = p_a + p_b + p_c + p_d = 1$,
it can be shown that when the four nucleotides
have equal probability $1/4$,
the average number of flow cycles reaches its maximum,
while its variance reaches its minimum. 
In other words, for sequences
of a given length, on average it requires more flow cycles
to determine the sequences with equal nucleotide probability
than the sequences with unequal nucleotide probabilities,
but the variance of the number of flow cycles 
for sequences with equal nucleotide probability
is smaller.

\begin{figure}
  \centering
  \includegraphics[angle=270,width=\columnwidth]{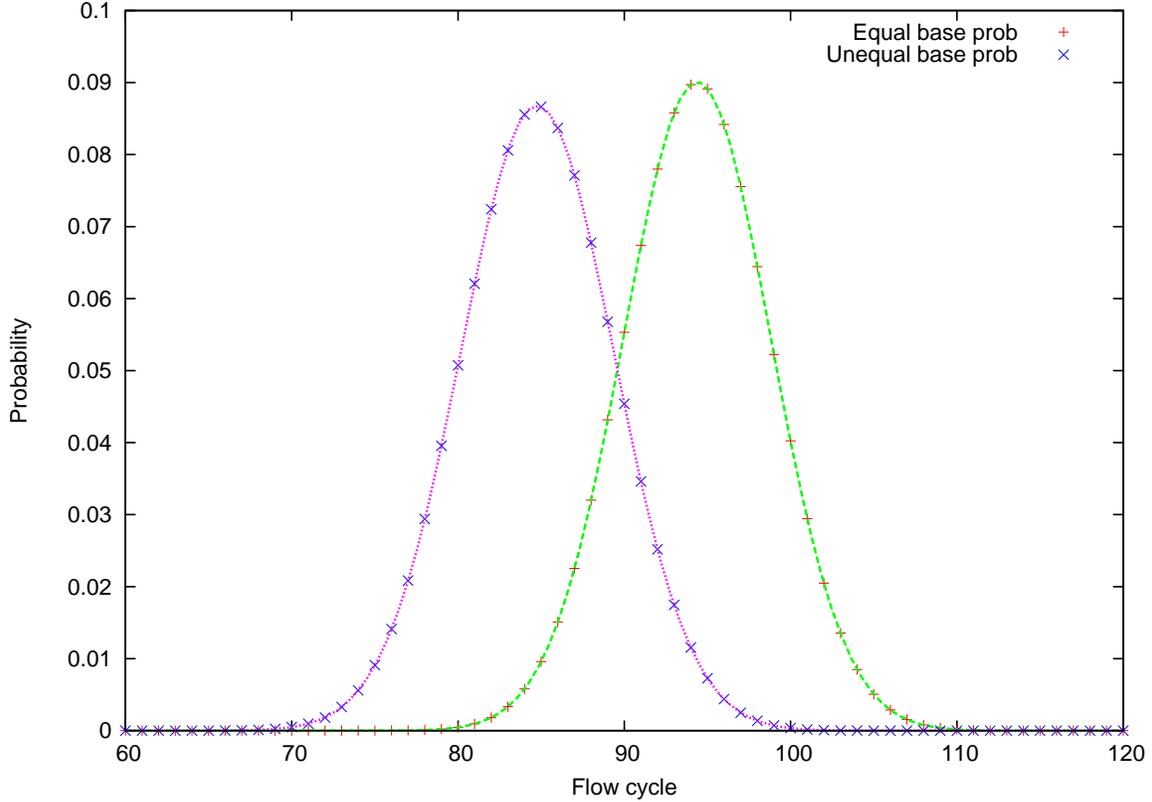}
  \caption{
    The distributions of flow cycles for a fixed sequence
    length of $n=250$ base pairs, for both equal nucleotide probability
    (on the right) and unequal nucleotide probabilities (on the left).
    The unequal nucleotide probabilities used here are
    $p_a=1/3=0.3333$, $p_b=1/11=0.0909$, $p_c = 100/231=0.4329$,
    and $p_d=1/7=0.1429$.
    The exact distributions are calculated from Eq. (\ref{E:G_sum}).
    The continuous curves are 
    the normal distributions $N({\bar f} (n), \var_f^2 (n))$
    of the same mean and variance as those of the exact distributions,
    where ${\bar f} (n)$ and $\var_f^2 (n)$ are calculated
    from Eqs. (\ref{E:fixed_length_avg}) and (\ref{E:fixed_length_var}).
    The two normal distributions shown here are
    $N(94.375, 19.609375)$ and $N(84.765278, 21.121065)$, 
    for equal and unequal nucleotide probabilities,
    respectively.
    \label{F:fixed_sequence_length}} 
\end{figure}

In Figure~\ref{F:fixed_sequence_length}
the exact distributions of flow cycles are shown for a fixed sequence
length of $n=250$ base pairs, for both equal nucleotide probability
(on the right) and an artificial example of
unequal nucleotide probabilities (on the left).
The unequal nucleotide probabilities used here are
$p_a=1/3=0.3333$, $p_b=1/11=0.0909$, $p_c = 100/231=0.4329$,
and $p_d=1/7=0.1429$.
These exact distributions are calculated from Eq. (\ref{E:G_sum})
in the previous section.

Also shown in Figure~\ref{F:fixed_sequence_length} 
in continuous curves
are the 
normal distributions $N({\bar f} (n), \var_f^2 (n))$,
the mean ${\bar f} (n)$ and variance $\var_f^2 (n)$ of which are calculated
from Eqs. (\ref{E:fixed_length_avg}) and (\ref{E:fixed_length_var}).
It is evident that the exact distributions can be approximated
accurately by normal distributions
with the same mean and variance.
For our two examples here,
the normal distributions are 
$N(94.375, 19.609375)$ and $N(84.765278, 21.121065)$, 
for equal and unequal nucleotide probabilities,
respectively.


\subsection{Fixed flow cycle: distribution of sequence length} 
\label{ss:fixed_cycle}

When the number of flow cycles $f$ is fixed,
the mean ${\bar n}(f)$ 
and variance $\var_n^2 (f)$ of the length of the  sequences  that
can be determined by these flow cycles
can also be calculated
from the exact probability GFs as described
in the previous section, by using Eqs~\ref{E:avg_fixed_cycle} and 
\ref{E:var_fixed_cycle}:
\begin{align} 
 {\bar n} (f) &\approx \frac{f}{\s_2} + \frac{2 \s_3}{\s_2^2} - 2 ,
  \label{E:fixed_cycle_avg}  \\
 \var_n^2 (f) &\approx  \frac{\s_2 - 3 \s_2^2 + 2 \s_3}{\s_2^3} f
 - 2 \frac{3 s_2 s_4 - 4 s_3^2 + s_2^2 s_3}{s_2^4}  .
 \label{E:fixed_cycle_var} 
\end{align} 
As discussed in section~\ref{SS:norm},
the small extra terms are ignored in the above expressions.
Practically they do not affect any discussions in the following.

From Eqs. (\ref{E:fixed_cycle_avg}) and (\ref{E:fixed_cycle_var}) we can see
that both the average sequence length ${\bar n} (f)$ 
and the variance $\var_n^2 (f)$
increase linearly with the number of flow cycle $f$.
When $p_a = p_b = p_c = p_d = 1/4$, we have
\begin{align}
  {\bar n} (f) &\approx \frac{8}{3} f - \frac{10}{9}   ,
  \label{E:fixed_cycle_eq_avg}  \\
  \var_n^2 (f) &\approx \frac{40}{27} f - \frac{20}{81} .
  \label{E:fixed_cycle_eq_var} 
\end{align}
By using the constraint $s_1 = p_a + p_b + p_c + p_d = 1$,
it can be shown that when the four nucleotides have equal probability,
both of the average sequence length and the variance
are at their mimima.
In other words, for a given number of flow cycles, 
on average longer read lengths
can be achieved when the target sequences have unequal nucleotide probabilities,
but the variance is also greater than that of the sequences with
equal nucleotide probability.

\begin{figure} 
  \centering
  \includegraphics[angle=270,width=\columnwidth]{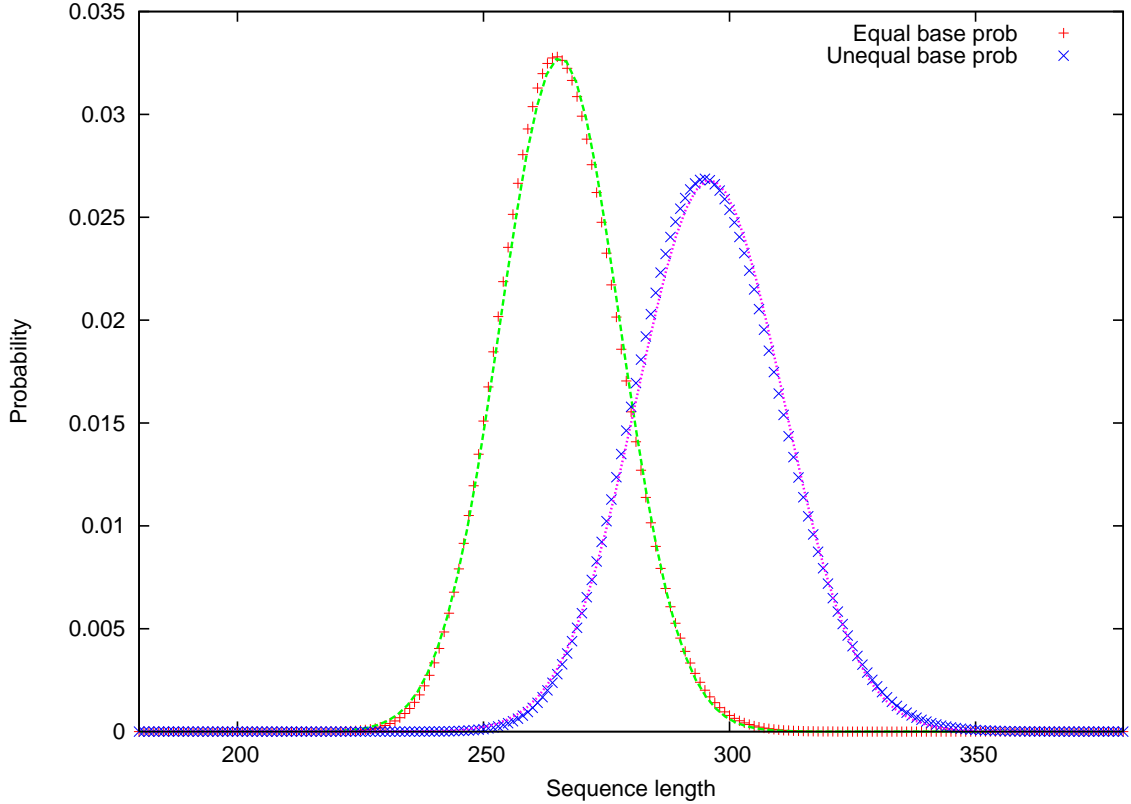}
  \caption{
    The distributions of sequence length in base pairs
    for a fixed number of flow cycles
    $f=100$, for both equal nucleotide probability
    (on the left) and unequal nucleotide probabilities (on the right).
    The unequal nucleotide probabilities used here are
    the same as in Figure~\ref{F:fixed_sequence_length}.
    The exact distributions are calculated from Eq. (\ref{E:G_sum}).
    The continuous curves are 
    the normal distributions $N({\bar n} (f), \var_n^2 (f))$
    with the same mean and variance as those of the exact distributions,
    where ${\bar n} (f)$ and $\var_n^2 (f)$ are calculated
    from Eqs. (\ref{E:fixed_cycle_avg}) and (\ref{E:fixed_cycle_var}).
    The two normal distributions  shown here are 
    $N(265.5555556, 148.3950617)$ and $N(296.0312085, 221.46233357)$, 
    for equal and unequal nucleotide probabilities,
    respectively.   
    \label{F:fixed_cycle}} 
\end{figure}

In Figure~\ref{F:fixed_cycle}
the exact distributions of sequence length in base pairs are shown for a fixed
number of flow cycles $f=100$ ($400$ nucleotide flows), 
for both equal nucleotide probability
(on the left) and unequal nucleotide probabilities (on the right).
The unequal nucleotide probabilities used here are the same as in the
previous section.
These exact distributions are calculated from  Eq. (\ref{E:G_sum})
in section~\ref{S:exact}.

Also shown in Figure~\ref{F:fixed_cycle} 
in continuous curves
are the 
normal distributions $N({\bar n} (f), \var_n^2 (f))$,
the mean ${\bar n} (f)$ and variance $\var_n^2 (f)$ of which are calculated
from Eqs. (\ref{E:fixed_cycle_avg}) and (\ref{E:fixed_cycle_var}).
Just like the distributions of the number of flow cycles at fixed sequence
length as discussed in the previous section,
the exact distributions of sequence length at a fixed  number of flow cycles
can also be approximated quite well with normal distributions
with the same mean and variance
as those of the exact distributions.
For the two examples here, the 
normal distributions are 
$N(265.5555556, 148.3950617)$ and $N(296.0312085, 221.46233357)$, 
for equal and unequal nucleotide probabilities,
respectively.

Compared with the almost perfect fit between the exact and normal
distributions in Figure~\ref{F:fixed_sequence_length},
the curves of normal distributions in Figure~\ref{F:fixed_cycle}
show some small disagreements with the exact distributions.
The difference here is that the exact distributions
have a slightly longer tails on the right and a slightly shorter
tail on the left when compared to the normal distributions.
Intuitively this discrepancies can be understood by the fact that
the distributions of sequence length
discussed in this section
include sequences of all theoretically possible lengths,
from the order of $f$ 
up to infinity. When the sequence length is fixed, as discussed in the previous
section, however,
there are only finite number of possible flow cycles.
For a sequence with length $n$, the number of nucleotide flows is from
$1$ to $3n+1$, as discussed previously in section~\ref{S:intro}.


\subsection{Fixed sequence length: 
distributions of individual nucleotide flows}
In the previous two sections we discussed
the distributions with regards to the flow cycles, 
which by definition are the collective behaviors of the flows
within the ``quad cycle'' (see the definitions in Table~\ref{T:defs}).
The advantage to deal with the quad flow cycles
instead of the individual nucleotide flows
is the symmetry in the expressions:
the master GF Eq.~(\ref{E:G_sum})
and the results derived from it (Eqs.~(\ref{E:fixed_length_avg}),
(\ref{E:fixed_length_var}), (\ref{E:fixed_cycle_avg}), 
and (\ref{E:fixed_cycle_var}))
are symmetric in the nucleotide probabilities
$p_a$, $p_b$, $p_c$, and $p_d$: these individual probabilities
do not appear in the expressions; only 
the elementary symmetric functions $s_i$'s are in the equations. 
If any two nucleotide probabilities are swapped, 
the results will not be affected.

In this and the next sections 
the distributions with respect to the individual nucleotide flows
will be presented.
Here we discuss explicitly the probabilities that nucleotide flows
ending in $a$, $b$, $c$, or $d$ will finish the sequence.
For example, the first sample sequence in Table~\ref{T:defs}
finishes at flow number $12$ at a nucleotide $d$, 
while the second sample sequence finishes at flow number $7$
at a nucleotide $c$.
As expected, the expressions will no longer be symmetric
with respect to nucleotide probabilities.
We'll first present the distributions of individual nucleotide flows
at fixed sequence length below.
They are derived from the GFs listed in
the section~\ref{S:exact}
(Eqs.~(\ref{E:G_a}), (\ref{E:G_b}), (\ref{E:G_c}), and (\ref{E:G_d})).
The averages of the  number of flow cycles that is required
to determine sequences of a length of $n$ 
with the last nucleotide flow
ending in each of the four different nucleotides are:
\begin{align} \label{E:fixed_length_ind_avg}
{\bar f_a} (n) &= \s_2 n    + 1-2 \s_2+p_b+p_c+p_d , \notag \\
{\bar f_b} (n) &= \s_2 n    + 1-2 \s_2+p_c+p_d     , \notag \\
{\bar f_c} (n) &= \s_2 n    + 1-2 \s_2+p_d         , \notag \\
{\bar f_d} (n) &= \s_2 n    + 1-2 \s_2             , 
\end{align}
and the variances
\begin{align} \label{E:fixed_length_ind_var}
\var_{f_a}^2 &= (2\s_3-3 \s_2^2+\s_2) n
               +8 \s_2^2 -6 \s_3 -2 \s_2 
	       + \sum_{i=b,c,d} ( p_i - p_i^2 - 2 p_i s_2  )    , \notag  \\
\var_{f_b}^2 &= (2 \s_3-3 \s_2^2+\s_2) n
               +8 \s_2^2 -6 \s_3 -2 \s_2
	       + \sum_{i=c,d} ( p_i - p_i^2 - 2 p_i s_2  )       , \notag \\
\var_{f_c}^2 &= (2 \s_3-3 \s_2^2+\s_2) n
               +8 \s_2^2 -6 \s_3 -2 \s_2
               +p_d-p_d^2-2 p_d \s_2                            ,  \notag \\
\var_{f_d}^2 &= (2 \s_3-3 \s_2^2+\s_2) n
               +8 \s_2^2-6 \s_3-2 \s_2                           . 
\end{align}
From these expressions we see again that the mean and
variance of the distributions vary linearly with the
sequence length $n$.

\begin{figure}
  \centering
  \includegraphics[angle=270,width=\columnwidth]{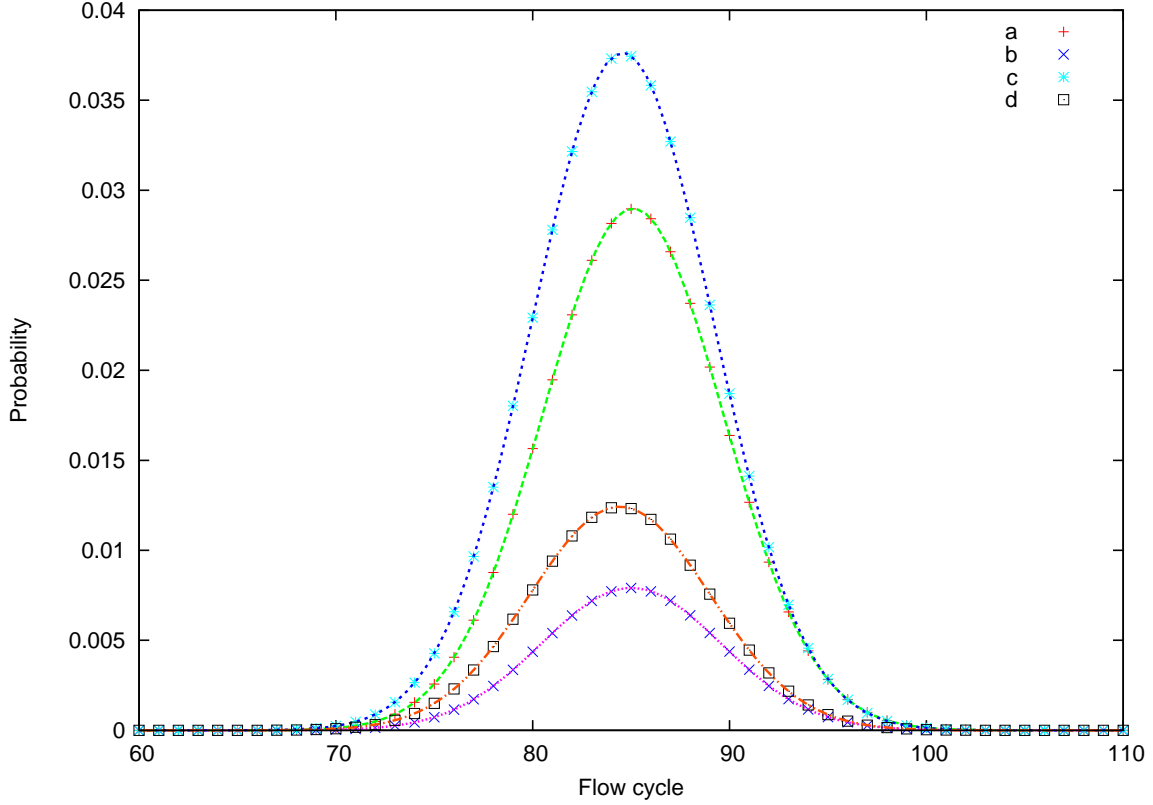}
  \caption{
    The distributions of individual flow cycles for a fixed sequence
    length of $n=250$ base pairs.
    The nucleotide probabilities used here are
    the same as  those in the unequal probability case
    in Figure~\ref{F:fixed_sequence_length}.
    These exact distributions are calculated from
    Eqs. (\ref{E:G_a}), (\ref{E:G_b}), (\ref{E:G_c}), and (\ref{E:G_d})
    in section~\ref{S:exact}.
    The continuous curves are 
    the normal distributions $N({\bar f_i} (n), \var_i^2 (n))$
    with the same means ${\bar f_i} (n)$ 
    and variances $\var_{f_i}^2 (n)$ as those of the exact distributions,
    which are calculated
    from Eqs. (\ref{E:fixed_length_ind_avg}) 
    and (\ref{E:fixed_length_ind_var}) for $i=a,b,c,d$.
    The normal distributions are scaled by the normalization factors
    of Eq. (\ref{E:normal_x}).
  \label{F:fixed_sequence_length_individual}} 
\end{figure}

In Figure~\ref{F:fixed_sequence_length_individual}
the exact distributions of flow cycles are shown for the four
nucleotides at a fixed sequence length of $n=250$ for unequal
nucleotide probabilities in the target sequences.
The nucleotide probabilities used here are the same as in the
previous sections.
These exact distributions are calculated from
Eqs. (\ref{E:G_a}), (\ref{E:G_b}), (\ref{E:G_c}), and (\ref{E:G_d})
in section~\ref{S:exact}.
We can see that the magnitudes of the probabilities in the
distributions are proportional to
the nucleotide frequencies.
From $a$ to $d$, the mean shifts to the left slightly, 
which can be confirmed by Eq. (\ref{E:fixed_length_ind_avg}).

Also shown in Figure~\ref{F:fixed_sequence_length_individual} 
in continuous curves
are the 
normal distributions $N({\bar f_i} (n), \var_{f_i}^2 (n))$,
for $i=a,b,c,d$,
with the same means and variances as those of the exact distributions,
calculated
from Eqs. (\ref{E:fixed_length_ind_avg}) and (\ref{E:fixed_length_ind_var}).
The normal distributions are multiplied by the normalization factors
Eq. (\ref{E:normal_x})
discussed in section~\ref{S:exact}.
As for the fixed sequence length case with quad cycles discussed previously 
(Figure~\ref{F:fixed_sequence_length}),
the approximations by normal distributions are almost perfect.


\subsection{Fixed flow cycle: 
distributions of sequence length ending with a particular individual 
nucleotide flow }
In this section the mean and variance of sequence length
that can be determined at a fixed number of flow cycles
with the last nucleotide flow ending at one of the
four different nucleotides are given.
For the means, we have
\begin{align} \label{E:fixed_cycle_ind_avg}
{\bar n_a} (f)   &\approx   \frac{1}{\s_2} f
         -\frac{(p_b + p_c + p_d) \s_2 - 2 \s_3 + \s_2^2}{\s_2^2} ,\notag \\
{\bar n_b} (f)   &\approx   \frac{1}{\s_2} f
         -\frac{(p_c + p_d) \s_2 - 2 \s_3 + \s_2^2}{\s_2^2}       ,\notag \\
{\bar n_c} (f)   &\approx   \frac{1}{\s_2} f
         -\frac{p_d \s_2 - 2 \s_3 + \s_2^2}{\s_2^2}               ,\notag \\
{\bar n_d} (f)   &\approx   \frac{1}{\s_2} f
         -\frac{- 2 \s_3 + \s_2^2}{\s_2^2}                       , 
\end{align}
and for the variances,
\begin{align}\label{E:fixed_cycle_ind_var}
\var^2_{n_a}     &\approx  \frac{2 \s_3-3 \s_2^2+\s_2}{\s_2^3}  f 
-\frac{-8 \s_3^2 + 6 \s_4 \s_2 + 2 \s_2^2 \s_3 
   + {\displaystyle \sum_{i=b,c,d} 
     ( p_i^2 s_2^2 - p_i s_2^3 + 2 p_i s_2 s_3 ) } }
{\s_2^4}                                                      ,  \notag   \\
\var^2_{n_b}     &\approx  \frac{2 \s_3-3 \s_2^2+\s_2}{\s_2^3}  f 
-\frac{-8 \s_3^2 + 6 \s_4 \s_2 + 2 \s_2^2 \s_3 
   + {\displaystyle \sum_{i=c,d} 
     ( p_i^2 s_2^2 - p_i s_2^3 + 2 p_i s_2 s_3 ) } }
   {\s_2^4}                                                   ,  \notag \\
\var^2_{n_c}     &\approx  \frac{2 \s_3-3 \s_2^2+\s_2}{\s_2^3}  f 
-\frac{-8 \s_3^2 + 6 \s_4 \s_2 + 2 \s_2^2 \s_3 
  + p_d^2 \s_2^2  
  - p_d   \s_2^3  
  + 2 p_d \s_2  \s_3 } {\s_2^4}                                , \notag \\
\var^2_{n_d}     &\approx  \frac{2 \s_3-3 \s_2^2+\s_2}{\s_2^3}  f 
-2 \frac{-4 \s_3^2+3 \s_4 \s_2+\s_2^2 \s_3}{\s_2^4}           .  
\end{align}
As in section~\ref{ss:fixed_cycle} for the ``quad cycle''
distribution of sequence length,
the small extra terms in these expressions
are not shown here for clarity reason.

\begin{figure}
  \centering
  \includegraphics[angle=270,width=\columnwidth]{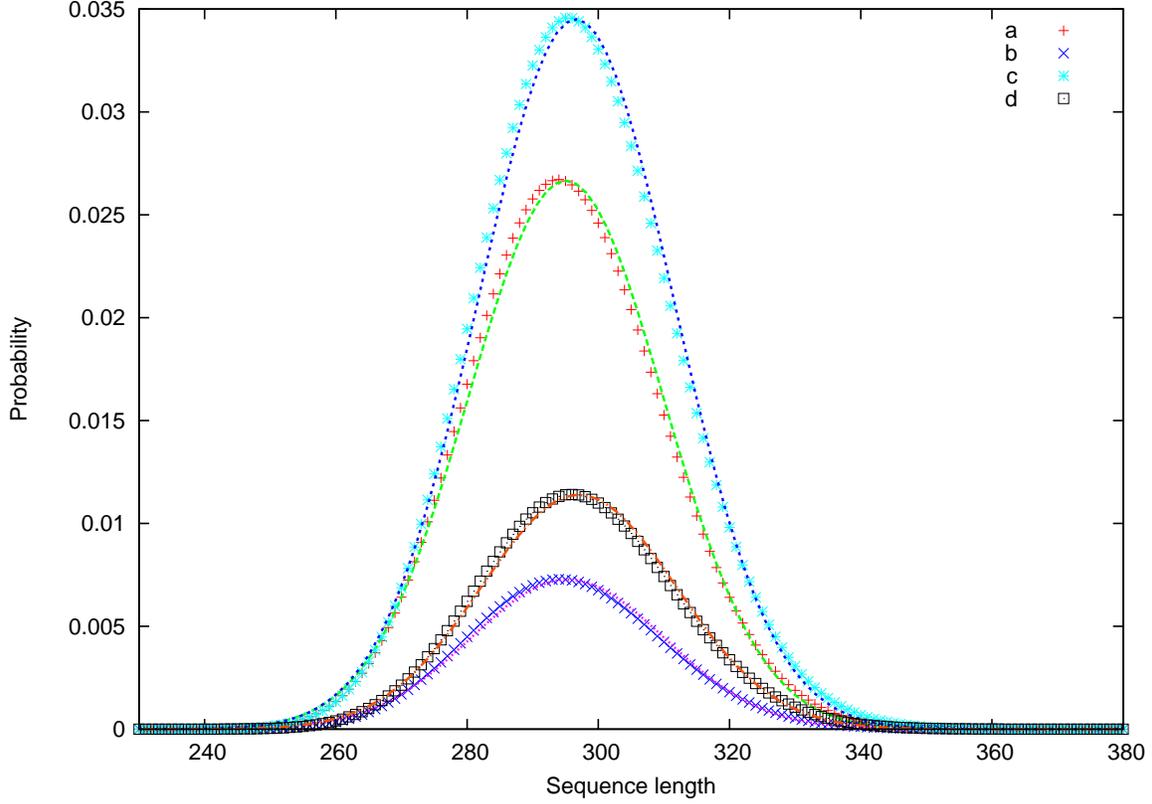}
  \caption{
    The distributions of of sequence length that can be determined
    with the number of flow cycles $f=100$, and with the last
    flow ending in the four different
    nucleotides.
    The nucleotide probabilities used here are
    the same as those in the unequal probability case
    in Figure~\ref{F:fixed_sequence_length}.
    These exact distributions are calculated from
    Eqs. (\ref{E:G_a}), (\ref{E:G_b}), (\ref{E:G_c}), and (\ref{E:G_d})
    in section~\ref{S:exact}.
    The continuous curves are 
    the normal distributions $N({\bar n_i} (f), \var_{n_i}^2 (f))$
    with the same means ${\bar n_i} (f)$ 
    and variances $\var_{n_i}^2 (f)$ as those of the exact distributions,
    which are calculated
    from Eqs. (\ref{E:fixed_cycle_ind_avg}) 
    and (\ref{E:fixed_cycle_ind_var}) for $i=a,b,c,d$.
    The normal distributions are scaled by the normalization factors
    of Eq. (\ref{E:normal_y}).
  \label{F:fixed_cycle_individual}} 
\end{figure}

In Figure~\ref{F:fixed_cycle_individual}
the exact distributions of the length of the 
sequences  that can be determined
with $f=100$  flow cycles and with the last
flow ending in one of the four different nucleotides are shown.
The nucleotide probabilities used here are the same as in the
previous sections.
These exact distributions are calculated from
Eqs. (\ref{E:G_a}), (\ref{E:G_b}), (\ref{E:G_c}), and (\ref{E:G_d})
in section~\ref{S:exact}.

Also shown in Figure~\ref{F:fixed_cycle_individual} 
in continuous curves
are the 
normal distributions $N({\bar n_i} (f), \var_{n_i}^2 (f))$,
$i=a,b,c,d$,
with the same means and variances 
as those of the exact distributions,
calculated
from Eqs. (\ref{E:fixed_cycle_ind_avg}) and (\ref{E:fixed_cycle_ind_var}).
The normal distributions are multiplied by the normalization factors
Eq. (\ref{E:normal_y})
discussed in section~\ref{S:exact}.
Similar to the quad cycles case with fixed number of cycles
discussed previously,
the sequence length distributions can be 
approximated accurately by the normal distributions
with the same mean and variance,
with slightly longer tails on the right and a slightly short tails 
on the left for the exact distributions.
From $a$ to $d$, the mean shifts to the right slightly,
which can be confirmed by Eq.~(\ref{E:fixed_cycle_ind_avg}).


\section{Discussion}
We have derived the statistical distributions for the pyrosequencing technique,
one of the important platforms for the next-generation sequencing technology.
Two cases are considered: 
the distribution of the number of flow cycles 
for fixed sequence length and the distribution of 
sequence length for fixed number
of flow cycles.
In both cases we gave explicit formulas for
the mean and variance of the distributions.
We  demonstrated that 
these distributions can be approximated accurately
by normal distributions with the same
  mean and variance calculated by the explicit formulas.
Both means and variance vary linearly with the number of flow cycles
(or sequence length).
When the  number of flow cycles is fixed,
the unequal frequencies of the nucleotides in the target sequences
will lead to longer read length with bigger variance 
(Figure~\ref{F:fixed_cycle});
Reciprocally, 
for a fixed sequence length it will need fewer flow cycles 
to determine the sequences if the frequencies of 
the nucleotides in the target sequences are unequal 
(Figure~\ref{F:fixed_sequence_length}).
These explicit formulas and qualitative statements
will be useful for instrument and software development
of the pyrosequencing platform,
and can be used as a guide in monitoring the machine performance
for daily users.



\end{document}